**Using atom probe tomography to understand Schottky barrier height pinning at the ZnO:Al / SiO2 / Si interface**

R. Jaramillo[1*], Amanda Youssef[2], Austin Akey[2], Frank Schoofs[3], Shriram Ramanathan[3,4], Tonio Buonassisi[2]

1. Department of Materials Science and Engineering, Massachusetts Institute of Technology, 77 Massachusetts Ave., Cambridge, MA 02139, USA

2. Department of Mechanical Engineering, Massachusetts Institute of Technology, 77 Massachusetts Ave., Cambridge, MA 02139, USA

3. School of Engineering and Applied Sciences, Harvard University, 9 Oxford St., Cambridge, MA 02138, USA

4. School of Materials Engineering, Purdue University, West Lafayette, IN 47907, USA

* rjaramil@mit.edu

*Abstract*

We use electronic transport and atom probe tomography to study ZnO:Al / $SiO_2$ / Si Schottky junctions on lightly-doped *n*- and *p*-type Si. We vary the carrier concentration in the the ZnO:Al films by two orders of magnitude but the Schottky barrier height remains constant, consistent with Fermi level pinning seen in metal / Si junctions. Atom probe tomography shows that Al segregates to the interface, so that the ZnO:Al at the junction is likely to be metallic even when the bulk of the ZnO:Al film is semiconducting. We hypothesize that Fermi level pinning is connected to the insulator-metal transition in doped ZnO, and that controlling this transition may be key to un-pinning the Fermi level in oxide / Si Schottky junctions.

*Text*

The electron energy band alignment at the interface between oxides and Si is a critical aspect of several technologies. Examples include the development of alternative gate dielectrics for continued field effect transistor scaling, and the development of carrier-selective contacts with wide bandgaps for Si heterojunction solar cells.[1–4] All such applications would benefit from controlling the band offset at the oxide / Si interface. Here we consider what happens when the oxide is electrically conductive. For the interface between conventional metals and Si the band alignment is weakly dependent on the metal work function.[5] Schottky barrier height measurements suggest that the Fermi energy ($E_F$) is pinned in the lower half of the Si bandgap, and several mechanisms have been proposed to explain these observations.[6] Will Fermi level pinning still occur when the metal is replaced by a conductive oxide? More significantly, can the band alignment at oxide / Si heterojunctions be tuned by controlling the conductivity of the oxide? We show that for ZnO the band alignment with Si can be changed by doping with Al, but also that control over this process is inhibited by chemical segregation at the interface. Material heterogeneity governs device performance and presents an obstacle to device engineering. Our results highlight the need for advanced microscopy to interpret device results, and suggest a strategy for controlling the band alignment at the interface between conducting oxides and Si.

Al-doped ZnO (AZO) is an important transparent conductor for solar cells, lighting, and transparent electronics. Like most oxides, AZO offers a large synthesis-properties-performance parameter space, and key properties such as work function, carrier concentration, mobility, and optical bandgap can be controlled by the film synthesis conditions.[7] The work function of AZO and other ZnO-based transparent conductors has been reported to vary from 4.2 to 5.1 eV.[7–9] Taken at face value, this implies that the Schottky barrier height ($\Phi$) at an AZO-semiconductor junction could be tuned by 0.9 eV. However, most reported work function measurements were performed on the exposed surface of AZO films, and therefore are not relevant to a buried junction. Moreover, the Schottky barrier height at metal-Si heterojunctions is only weakly dependent on the metal work function.

We synthesized AZO / $SiO_2$ / Si Schottky diodes by reactive sputtering of AZO films on Si substrates. We controlled the AZO film properties by changing the oxygen gas flow into the growth chamber. We use the transparent conductor figure of merit (FOM) to define a convenient metric to define the oxygen content in the resulting films.[8] Oxygen-poor films are conductive but not transparent, and oxygen-rich films are transparent but not conductive. The oxygen content is measured as a percentage of oxygen flow relative to the optimal film, and we write this relative oxygen content as $R_{O2}$. Hall measurements show that the free electron concentration at room temperature varies from $(8.2 \pm 0.3) \times 10^{20}$ cm$^{-3}$ for oxygen-poor films to $(1.14 \pm 0.05) \times 10^{19}$ cm$^{-3}$ for oxygen-rich films. The insulator-metal transition for ZnO is in the range of 5 - 8 x $10^{19}$ cm$^{-3}$, meaning that our study includes semiconducting and metallic samples.[10] We deposited AZO simultaneously on *n*- and *p*-type Si substrates with carrier concentrations of $(4.1 \pm 1.5) \times 10^{15}$ cm$^{-3}$ and $(2.2 \pm 0.5) \times 10^{15}$ cm$^{-3}$, respectively, as measured by Mott-Schottky capacitance profiling. Before AZO deposition the substrates were prepared by forming a controlled $SiO_2$ native oxide layer, in order to avoid uncontrolled $SiO_2$ formation during the AZO sputtering process. The $SiO_2$ layer is $15 \pm 2$ Å thick and we presume that it is easily tunneled through. Please see the Supporting Information (SI) for full details of our sample preparation, including sputtering conditions, the preparation of ohmic contacts to the substrates, and the characterization of the $SiO_2$ layer.[11]

We determined $\Phi_b$ by measuring current-voltage ($J(V)$) curves as a function of temperature ($T$), and analyzing the reverse saturation current density ($J_0(T)$) using the standard model of thermionic emission.[11]

We find that $\Phi$ is independent of AZO film properties for both *n*- and *p*-type diodes. We plot the AZO carrier concentration in **Figure 1a** and the measured barrier heights in **Figure 1b**. It seems that $E_F$ is pinned at $\Phi_p = 0.37 \pm 0.03$ eV above the valence band edge ($E_V$) for *p*-type diodes, and $\Phi_n = 0.53 \pm 0.02$ eV below the conduction band edge ($E_C$) for *n*-type diodes. However, these values are under-estimated because our samples are M-I-S diodes, and our analysis does not account for barrier height lowering. This is clear from the sum $\Phi_p + \Phi_n = 0.90 \pm 0.05$ eV, which is lower than the Si bandgap ($E_g^{Si} = 1.1$ eV). If we approximate that $\Phi$ is lowered equally for *n*- and *p*-type diodes then $E_F$ is pinned 0.47 eV above $E_V$ for *p*-type samples, and 0.63 eV below $E_C$ for *n*-type samples.

In **Figure 2** we compare our estimate ($\Phi_p - \Phi_n$)/2 = -0.084 eV to the expectations for Fermi level pinning at the charge neutrality level (CNL), for an unpinned Fermi level ("electron affinity" alignment), and to previously published results on metal / Si and ZnO / Si diodes.[5,12–15] With the sole exception of Ca, AZO / Si and metal / Si diodes all exhibit $\Phi_n > \Phi_p$, meaning that the Fermi level falls in the lower half of the Si bandgap. This is qualitatively consistent with $E_F$ pinning at the CNL, 0.36 eV above $E_V$.[12] We also show the value expected if the Fermi level is unpinned, meaning that $\Phi_p$ and $\Phi_n$ are determined by the work function of ZnO (taken here as 4.5 eV).[11] In this case the ordering is reversed, $\Phi_n < \Phi_p$. There are several published reports of ZnO / Si diodes for which barrier height measurements were made on both *n*- and *p*-type substrates, and these results are reproduced in Figure 2. The ZnO / Si diode results cover both $\Phi_n$

> $\Phi_p$ and $\Phi_n$ < $\Phi_p$. The band alignment for sputtered, undoped ZnO closely matches the prediction for an unpinned Fermi level.[13] The difference in barrier height between our AZO / Si diodes and previously published ZnO / Si diodes is up to 0.2 eV. We conclude that the band alignment between ZnO and Si can be tuned by doping, and that for highly doped AZO the band alignment resembles that of conventional metals. We hypothesize that some mechanism creates a dipole at the AZO / Si interface that is not present at the ZnO / Si interface, and that this mechanism is connected to the electrical conductivity of AZO. However, the carrier concentration and conductivity vary widely across our sample set, which includes samples on the semiconducting side of the insulator-metal transition, and yet the Fermi level remains pinned. To address this apparent contradiction we turn to microscopy.

Band offsets are determined by dipoles that are tightly confined to the interface. Therefore, to understand our device results we should study the atomic-scale structure and composition of our interfaces. For this we turn to atom probe tomography (APT). APT is a destructive microscopy technique that can provide three dimensional chemical composition analysis of the interface with atomic resolution. APT requires needle-shaped samples, from which atoms are progressively field evaporated by applying a high positive voltage to the tip surface along with a triggering pulse (either a high-voltage pulse or a shot from an ultrafast pulsed laser). These atoms are projected onto a position sensitive detector and are identified by time-of-flight mass spectrometry (see SI for mass spectrum).[11,16,17] We prepared six samples for APT analysis: one each *p*- and *n*-type for oxygen-poor ($R_{O2}$ = -24.5%), optimal ($R_{O2}$ = 0%), and oxygen-rich ($R_{O2}$ = 26.7%) samples. The APT samples were prepared from the same Schottky diodes that were used for electrical measurements by focused-ion-beam lift-out using a Zeiss NVision 40 Dual Beam system. The tips were sharpened by annular ion milling until a tip radius on the order of 50-80 nm and shank angles of approximately 15 degrees (half angle) were reached. The six tips were measured by laser pulsed APT using a LEAP 4000X HR with a pulse energy of 100 pJ (355 nm, 50 fs FWHM), a 100 kHz pulse rate and a base temperature of 40 K. We reconstruct the field evaporated tips in 3D by measuring the tip shank angle and radius from high resolution scanning electron micrographs, assuming the tip surface is spherical. Reconstruction artifacts are noticeable especially at the AZO / $SiO_2$ / Si interface. This is due to the fact that different phases at the junction may have significantly different evaporation fields, causing the image compression factor to deviate from that assumed in reconstruction. [18–20]

In **Figure 3** we show two dimensional contour lines of the three dimensional Al concentration data along the depth of the needle-shaped tip. There is a significant segregation of Al to the AZO / $SiO_2$ / Si interface for all samples. This suggests that at the interface the AZO is always highly doped, even for films that are semiconducting in the bulk. The overall Al concentration decreases with increasing oxygen content, as expected from point defect thermodynamics.[21] The Al concentration values in Figure 3 have a source of inaccuracy, which is that the oxygen content cannot be accurately measured (because $O_2^{2+}$ and $O^{1+}$ are indistinguishable). As a result, the (Zn+Al):O ratios are nearly 2:1 throughout the AZO films, and the Al concentrations in Figure 3 are inflated.[11] The APT data can be reduced to one dimension through use of a proximity histogram ("proxigram"), that is, by averaging over planes perpendicular to the interface, which reduces distortions caused by curvature artifacts in the reconstruction, where the interface is defined as the depth at which Si reaches 5 at. %.[11] We also scale the metal concentrations such that the ratio (Zn+Al):O is 1:1 in the midst of the AZO layer. We plot the resulting Al composition profiles in **Figure 4a**.

Conduction electrons in AZO come from $Al_{Zn}^{\bullet}$ ionized donors. The corresponding Brouwer approximation is that the only charged species are $Al_{Zn}^{\bullet}$ and conduction electrons, and that their concentrations are equal: $Al_{Zn} \leftrightarrow Al_{Zn}^{\bullet} + e'$, $[Al_{Zn}^{\bullet}] = [e']$. Our AZO films are all *n*-type, and $Al_{Zn}$ is the donor with the lowest formation energy in both the oxygen-poor and oxygen-rich limits.[21] Therefore, we assume that the same Brouwer approximation applies for our full sample set. It is likely that Al exists in forms other than $Al_{Zn}^{\bullet}$ in our films, and the distribution of Al in its different forms would be sample-dependent (due to dependencies on oxygen content). Therefore, we assume a simple relationship between the concentration of ionized donors and the total Al concentration through a film-dependent factor *f*: $[Al_{Zn}^{\bullet}] = f[Al]$. Hall measurements give us $[Al_{Zn}^{\bullet}]$ averaged over the whole film, and APT data gives us [Al] far from the interface. Therefore, we can determine *f* for each sample, and use it to calculate $[Al_{Zn}^{\bullet}]$ through the interface. We plot the results in **Figure 4b**. For the semiconducting oxygen-rich film, the apparent concentration of ionized donors at the interface is sufficiently high to cross the insulator-metal transition. The simplifying assumption that the factor *f* is constant for a given sample overlooks the fact that the ionized fraction $[Al_{Zn}^{\bullet}]/[Al]$ could change with the local environment, and therefore the curves in Figure 4b should be considered upper bounds on the local carrier concentration.

The APT data suggests that, right at the interface, all of our films are on the metallic side of the insulator-metal transition, and that this is responsible for pinning the Fermi level. The exact composition at the interface varies from film to film (see SI for full Al, Zn, Si and O profiles)[11], and yet $\Phi_p$ and $\Phi_n$ remains nearly constant within the experimental error. There are small trends in the data that indicate some tuning of the barrier height, but these trends are overwhelmed by the change in band alignment on going from ZnO to AZO. It seems that the proximity of a metallic layer to the Si is sufficient to pin the Fermi level.

Device transport results (Fig. 2) show that the band alignment of ZnO / Si diodes can be changed by doping, and atom probe results (Fig. 4) suggest that the Fermi level is pinned by an insulator-metal transition at the interface, even for films that are semiconducting on-average. Since our samples are M-I-S structures, there is no intimate contact between the metal (AZO) and the Si. The intrinsic layer is thin enough to allow electron tunneling between AZO and Si. Taken together, these results suggest that the Fermi level pinning is related to the evanescent decay of delocalized electron wavefunctions from the metal into the Si, often called metal-induced gap states. Evidence for metal-induced gap states contributing to Fermi level pinning has been shown recently for the Fe / GaAs system.[22] A thin layer of NiO has been found to partially un-pin the Fermi level at metal / NiO / Si Schottky junctions.[4] However, the NiO layer was 26 Å thick, which is significantly larger for evanescent decay than the $SiO_2$ layer used here. These results call for a detailed study of the effect of insulator thickness and composition on Fermi level pinning at M-I-Si Schottky junctions.

We suggest that the Schottky barrier height on Si could be tuned by fine control of the carrier concentration at an insulator-metal transition in the metallic layer. Such fine control is inhibited in AZO by the segregation of Al to the interface. Therefore, achieving a tunable barrier height in ZnO / Si junctions may depend on controlling the mechanism that causes defect segregation at the interface. Theory predicts that steric energy will cause Al dopants to segregate to the boundary of small, isolated ZnO crystals, but it is not clear whether this mechanism can explain our observations at a buried interface.[23] Kinetic and/or electrostatic control of the interface during AZO growth may enable the desired control of defect segregation and thereby the insulator-metal transition.

Our interpretation of the APT data poses several questions. The biggest questions come from the fact that we do not know the phases at the interface. The Si is covered with a layer of $SiO_2$. The thickness of this layer is 15 ± 2 Å as measured by ellipsometry (APT cannot accurately measure the thickness of this layer due to abrupt changes of the evaporation field at the interface).[20] However, in the nominally AZO layer there could be secondary phases present. The concentration of Al at the interface exceeds the solid solubility limit of 2% Al/Zn (equal to [Al] = 8.3 x $10^{20}$ $cm^{-3}$), suggesting the presence of Al-rich phases such as $Al_2O_3$ and Al:Si (meant to include pure Al).[24] The presence of a uniform Al:Si layer at the interface would be consistent with a uniform barrier height (Fig. 1b). However, it is unlikely that Al:Si would form because proximity to $SiO_2$ and the reactive oxide growth environment mean that $Al_2O_3$ is strongly favored thermodynamically over metallic Al:Si. The presence of $Al_2O_3$ at the interface could affect the measured $J_0$, $\Phi_n$, and $\Phi_p$, but probably would not affect the position of $E_F$ in the Si bandgap. Discontinuous $Al_2O_3$ at the interface would change the measured $J_0$ by reducing the actual area of the AZO / $SiO_2$ / Si junction, but this would not affect the $T$-dependent diode analysis. A continuous $Al_2O_3$ film on top of the $SiO_2$ layer would increase the thickness of the intrinsic layer in our M-I-S devices, which would lower both $\Phi_n$, and $\Phi_p$. In particular, this could be responsible for some of the observed sample-to-sample variation in $\Phi_n$ + $\Phi_p$. We note that sputtering is a non-equilibrium growth method, and the presence of Al above the solid solubility limit does not necessarily imply the existence of secondary phases. In addition to the phase information it would be helpful to measure directly the free carrier concentration at the interface, rather than relying on the composition as a proxy for carrier concentration. High resolution transmission electron microscopy combined with electron energy loss spectroscopy to measure the bulk plasmon energy would help to answer these questions.

In summary, we show that the Fermi level is pinned at AZO / SiO$_2$ / Si Schottky diodes, and we suggest that this pinning is connected to the insulator-metal transition in AZO. Control of the diode is inhibited by segregation of Al to the interface. We suggest that the ZnO / Si barrier height could be tuned by at least 0.2 eV by engineering point-defect distributions in order to control the insulator-metal transition at the interface. Our results highlight the importance of interface chemistry in determining the Schottky barrier height on Si, and the role of heterogeneity in governing device performance.[6] We emphasize the need to combine device transport with appropriate high resolution microscopy to better understand electronic heterojunctions.

**Acknowledgments**

We acknowledge Harry Tuller and Michael Campion for conversations and for sharing unpublished data. R. Jaramillo acknowledges the support of a Department of Energy (DOE) EERE Postdoctoral Research Award. This work was supported by Bay Area Photovoltaic Consortium under DOE contract No. DE-EE0004946. We acknowledge support from the Air Force Office of Scientific Research under contract No. FA9550-12-1-0189, and from the National Science Foundation under contract No. DMR-0952794. This work was performed in part at the Center for Nanoscale Systems (CNS), a member of the National Nanotechnology Infrastructure Network (NNIN), which is supported by the National Science Foundation under contract No. ECS-0335765.

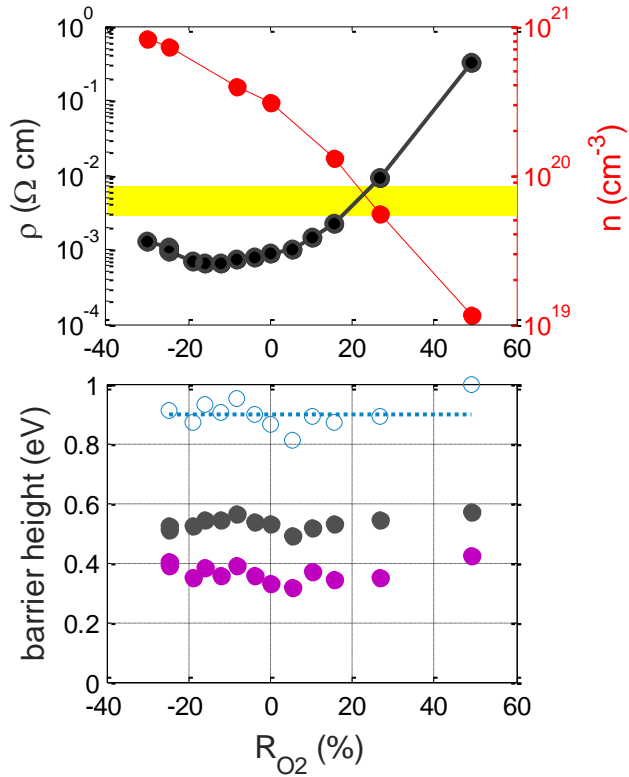

**Fig 1:** AZO film properties and Schottky barrier heights on *n*- and *p*-type Si. **(a)** Resistivity (left axis, black curve) and conduction electron concentration (right axis, red curve) for AZO films as a function of relative oxygen content $R_{O2}$ (see text). The yellow bar indicates the electron concentration corresponding to the insulator-metal transition in ZnO.[10] **(b)** Schottky barrier height for AZO / SiO$_2$ / Si diodes on both *n*-type (black points) and *p*-type (purple points) Si. The sum of barrier heights $\Phi_n + \Phi_p$ is shown in light blue. The fact that it is lower than the Si bandgap is likely due to barrier height lowering in our M-I-S diodes.

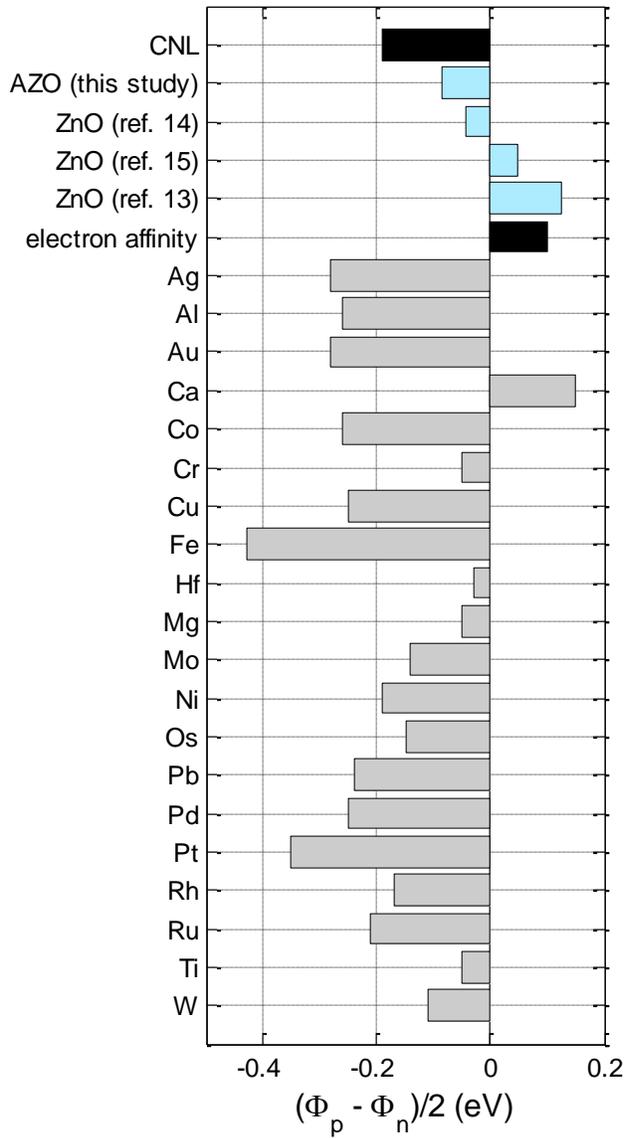

**Fig 2.** We compare the difference ($\Phi_p$ - $\Phi_n$)/2 found here to that reported in the literature for metals and ZnO (grown by sputtering and atomic layer deposition), and to predictions of barrier heights based on charge neutrality level (CNL) and electron affinity rules of band alignment.[5,12–15] See SI for details including the work function of ZnO the treatment of results in ref. 15.[11]

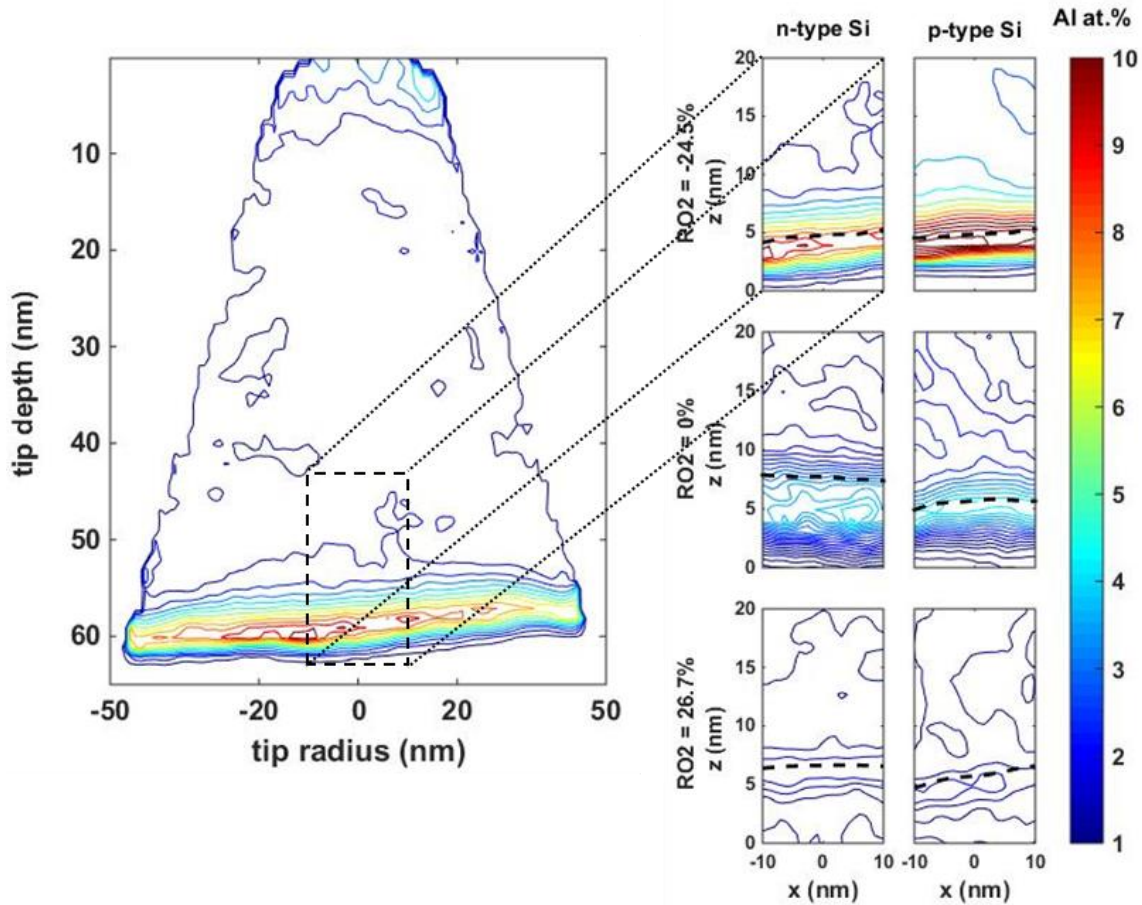

**Fig 3:** Al segregation at the AZO / SiO$_2$ / Si interface observed by APT. (**Left**) 2D contour plot of the Al concentration (at. %) for the oxygen poor ($R_{O2}$ = -24.5%) AZO layer on *n*-type Si substrate. The 2D projection from the entire tip volume contour plots was generated by sampling the 3D grid in the z-direction with a distance between the sampling planes equal to 1 nm. (**Right**) Panels showing 2D contour plots of the Al concentration for all 6 samples measured by APT. The dashed black line corresponds to the defined interface (5 at. % Si). 3D reconstruction artifacts such as curvature effects is caused by the difference in evaporation field needed to field evaporate the different elements present at the junction and the resulting change in image compression factor. Also, notice the interface line being tilted, this is the result of sample preparation wherein the tip is deposited slightly offset from its vertical position.

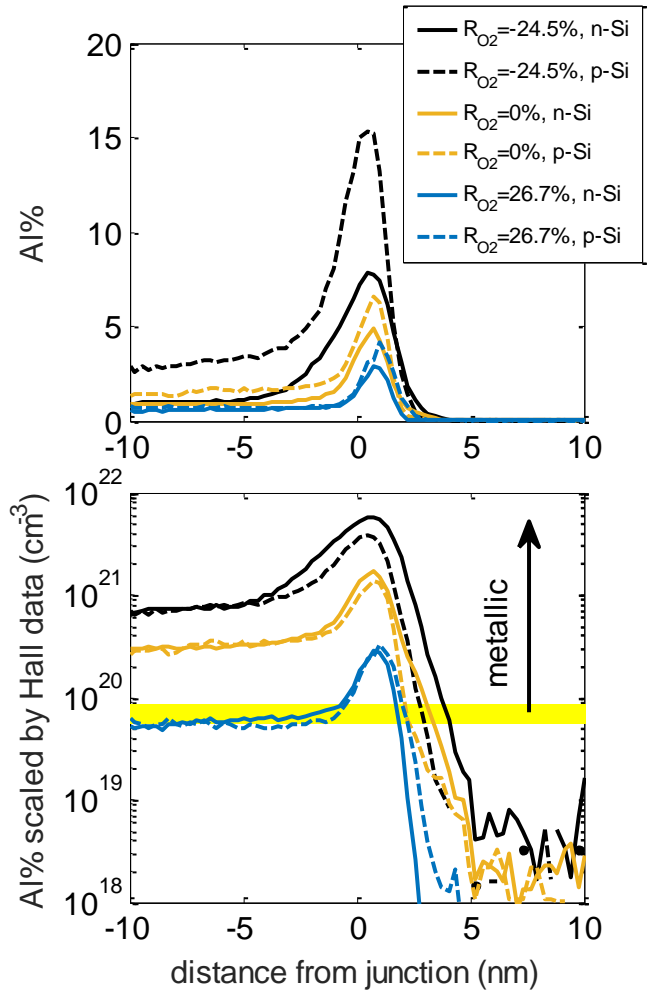

**Fig 4:** Al concentration across the AZO / SiO2 / Si interface, generated using a proxigram normal to an interface defined as 5 at. % Si. *n*-type samples are shown with solid lines, *p*-type are shown with dashed lines. Black: oxygen-poor ($R_{O2}$ = -24.5%). Orange: optimal ($R_{O2}$ = 0%). Blue: oxygen-rich ($R_{O2}$ = 26.7%). **(a)** Al at. %. **(b)** Al concentration scaled by the Hall data measured on witness AZO films. Yellow bar indicates the free electron concentration at the insulator-metal transition. The segregation of Al suggests that the concentration of ionized donors $Al_{Zn}^{\bullet}$ at the interface is always on the metallic side of the insulator-metal transition.

# Supplemental Material

1. AZO growth

    - We grew AZO films by reactive magnetron sputtering from a Zn:Al target, 1.2 wt. % Al (ACI Alloys, Inc.) using a sputtering system built by AJA International Inc.

    - The temperature of the substrate holder was maintained at 200 C during growth.

    - The total gas pressure was 2.8 mTorr, and the Ar gas flow was 39.7 sccm. The $O_2$ gas flow was varied from 3.9 to 10.1 sccm ($R_{O2}$ ranging from -36.0 to 49.3%).

    - The RF power to the target was 100 W

    - The total deposition time was constant at 1050 sec. The AZO film thickness varied between 204 and 118 nm, with the thickness decreasing with increasing $R_{O2}$ (more oxygen = slower growth). The thickness was measured by a combination of profilometry and ellipsometry.

2. AZO film properties

    - We show the figure of merit (FOM) and electrical properties of our films in **Figure S1**. The FOM is calculated by combining optical spectrophotometry data with electronic transport data. For both types of data, measurements were made on witness films grown on glass at the same time as our device samples.

    - We show X-ray diffraction data for our AZO films in **Figure S2**. These data were measured on witness films grown on oxidized Si wafers at the same time as our device samples. All samples are strongly c-axis textured.

    - We show optical spectroscopy data for our films in **Figure S3**. The absorption coefficient ($\alpha$) was determined from reflection ($R$) and transmission ($T$) data

measured on a spectrometer for films grown on glass, where multiple reflections in the sample are not accounted for.

- In Figures S1-S3, we show data for films with $R_{O2} \geq -36.0\%$. However, in the main text we only show Schottky devices results for $R_{O2} \geq -24.5\%$. The reason is that the most oxygen-poor films did not yield usable Schottky device data. We speculate that this is because the materials were thermodynamically unstable due to the oxygen deficiency, and were changing over time during handling and measurement.

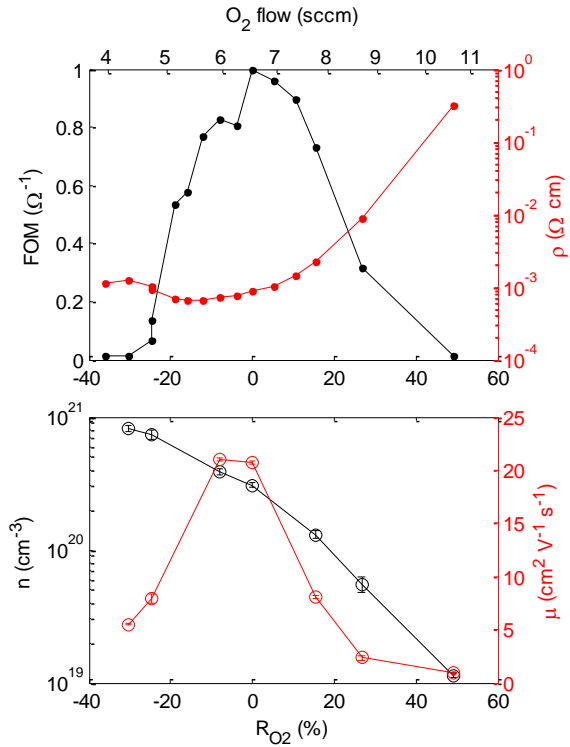

**Fig S1:** Electrical properties and FOM of AZO films as a function of $R_{O2}$. The top axis show the equivalent $O_2$ flow rates used during growth. (**Top**) FOM (black, left axis) and resistivity ($\rho$, red, right axis). The FOM is defined as in Gordon, using the window 400 – 1100 nm.[8] (**Bottom**)

Room-$T$ Hall results for the free electron concentration ($n$, black, left axis) and mobility ($\mu$, red, right axis).

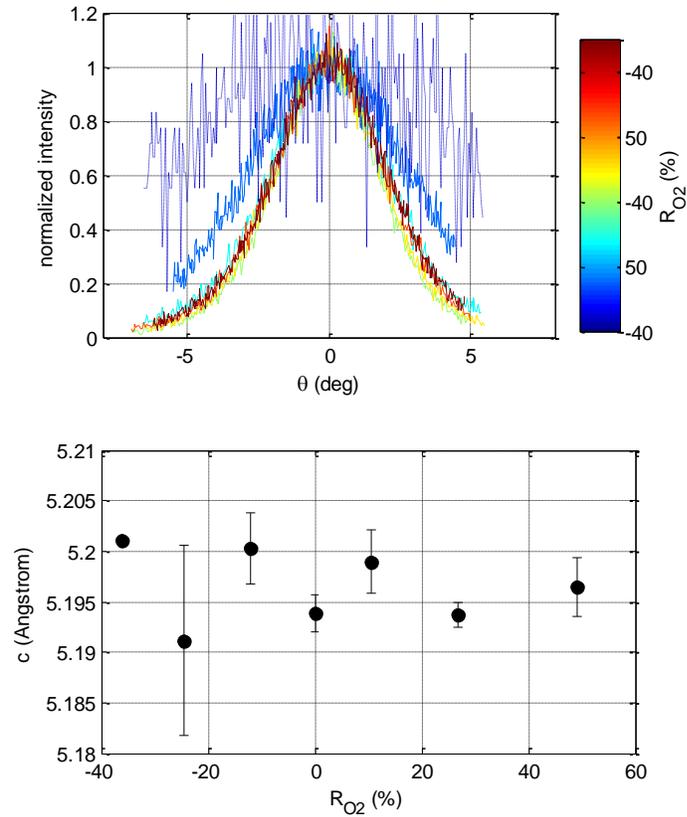

**Fig S2:** X-ray diffraction measurements on select AZO films. All samples were c-axis textured, and data is from symmetrical θ-2θ scans. (**Top**) Rocking curves for the (002) reflection as a function of $R_{O2}$, indicated by the color scale. The x-axis scale is set to zero at the peak. All films show similar texture and crystallinity except for the most oxygen-poor sample, $R_{O2} = -36.0\%$, which is shown with a dotted line. Schottky diodes made from this film did not produce usable data and are not included in this study. (**Bottom**) c-axis lattice constant as a function of $R_{O2}$.

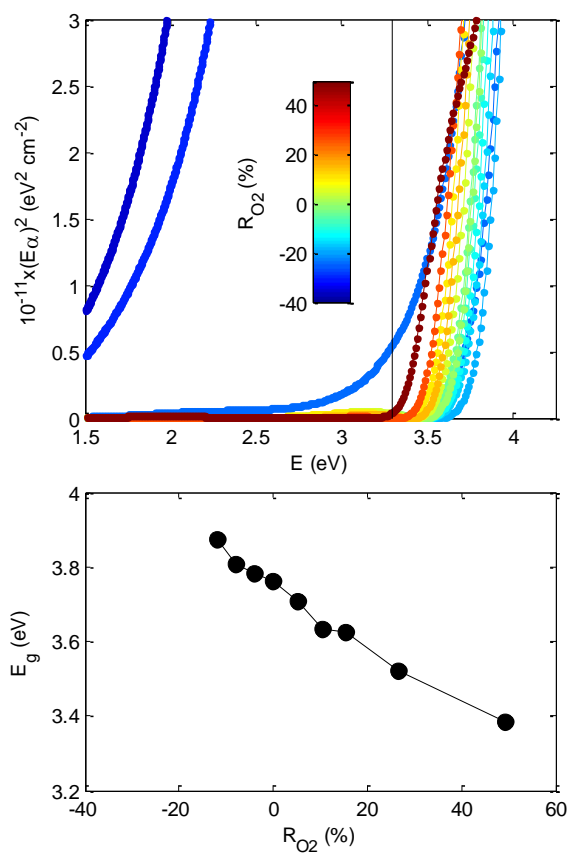

**Fig S3:** Optical spectroscopy of AZO films. (**Top**) Tauc plot appropriate for a direct bandgap material. For $R_{O2} \geq -12\%$ the data are well fit by a straight line, and the bandgap can be determined by the intercept of this line with the x-axis. For more oxygen-poor samples ($R_{O2} < -12\%$) there is strong sub-bandgap absorption, and the optical bandgap cannot be accurately determined from these data. The black vertical line marks the bandgap of undoped ZnO, $E_g = 3.3$ eV. Notice that the bandgap of AZO approaches this value in the oxygen-rich limit. (**Bottom**) Bandgap extracted from the Tauc plots.

3. Si / SiO$_2$ / AZO Schottky diode fabrication
    - We schematically illustrate our Schottky diode fabrication process in **Figure S4**.
    - We start with lightly doped *n*- and *p*-type Si wafers (phosphorous and boron doping, respectively). All wafers were specified with 1-10 Ω cm resistivity and (100) orientation.
    - Next we make ohmic contact to the wafers, over a large area on the back and at select contact areas on the front. For *n*-type wafers we use a spin-on dopant (SOD) to create highly doped n++ layers (P509, Filmtronics). The dopant is patterned on the front to define contact pads using photolithography. The SOD is diffused using a 950 C, 50 min anneal in an oxidizing environment. For this step it is important to have a diffusion barrier covering all areas on the front of the wafer that are not intended as ohmic contacts. Otherwise we found that dopant is transported through the environment and results in unwanted doping where SOD was not present.
    - For ohmic contact to p-type wafers we deposit Al by sputtering, followed by a rapid thermal anneal at 850 C for 100 s in an inert atmosphere, in order to create highly doped p++ layers.
    - Following the formation of n++ or p++ layers, these regions are metallized by depositing Ti / Au contact pads. The resulting I(V) curves measured between pairs of ohmic contacts are linear at all scales of our instrument.
    - The *n*- and *p*-type wafers with ohmic contacts are then prepared for AZO deposition. Due to the reactive growth environment and the thermodynamics of the ZnO / Si interface, a SiO$_2$ interface layer is unavoidable. We therefore chose

to control the process by intentionally oxidizing the Si wafers before AZO growth. We do this by first removing the native oxide by etching in a buffered HF solution, followed by a 5 min re-oxidation step in a UV/ozone chamber. Note that the metallized areas are protected by photoresist during the buffered HF dip.

- The wafers are then transferred to the sputtering chamber for AZO growth. AZO is grown as a blanket film.
- The wafers are then covered by a blanket film of Ti / Au for top electrical contact.
- The wafers are subsequently patterned into Schottky diode mesas using photolithography and wet etching. The data in this paper was measured on diodes of size 200 x 200 μm$^2$.
- The thickness of the interfacial $SiO_2$ layer was measured by spectroscopy ellipsometry. In **Figure S5** we show the results of these measurements for test samples that were carried through the device fabrication procedure. The as-delivered wafer has a $SiO_2$ thickness of approximately 4 nm. Following the buffered HF dip and UV/ozone re-oxidation, the $SiO_2$ thickness is approximately 1.5 nm thick. We also measure this thickness after AZO deposition, both by measuring a full AZO / $SiO_2$ / Si stack and by removing the AZO layer with a dilute HCl etch and measuring the resulting $SiO_2$ / Si stack. All measurements are consistent with a $SiO_2$ layer thickness of 15 ± 2 Å.

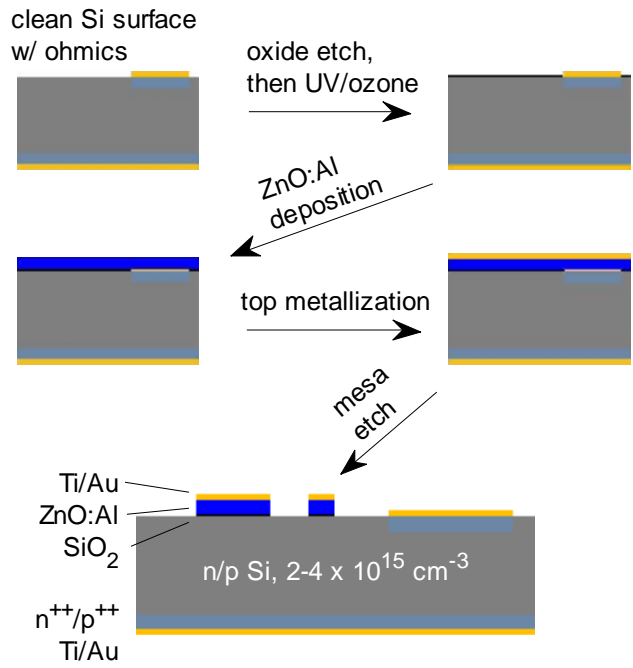

**Fig. S4:** Schottky diode fabrication.

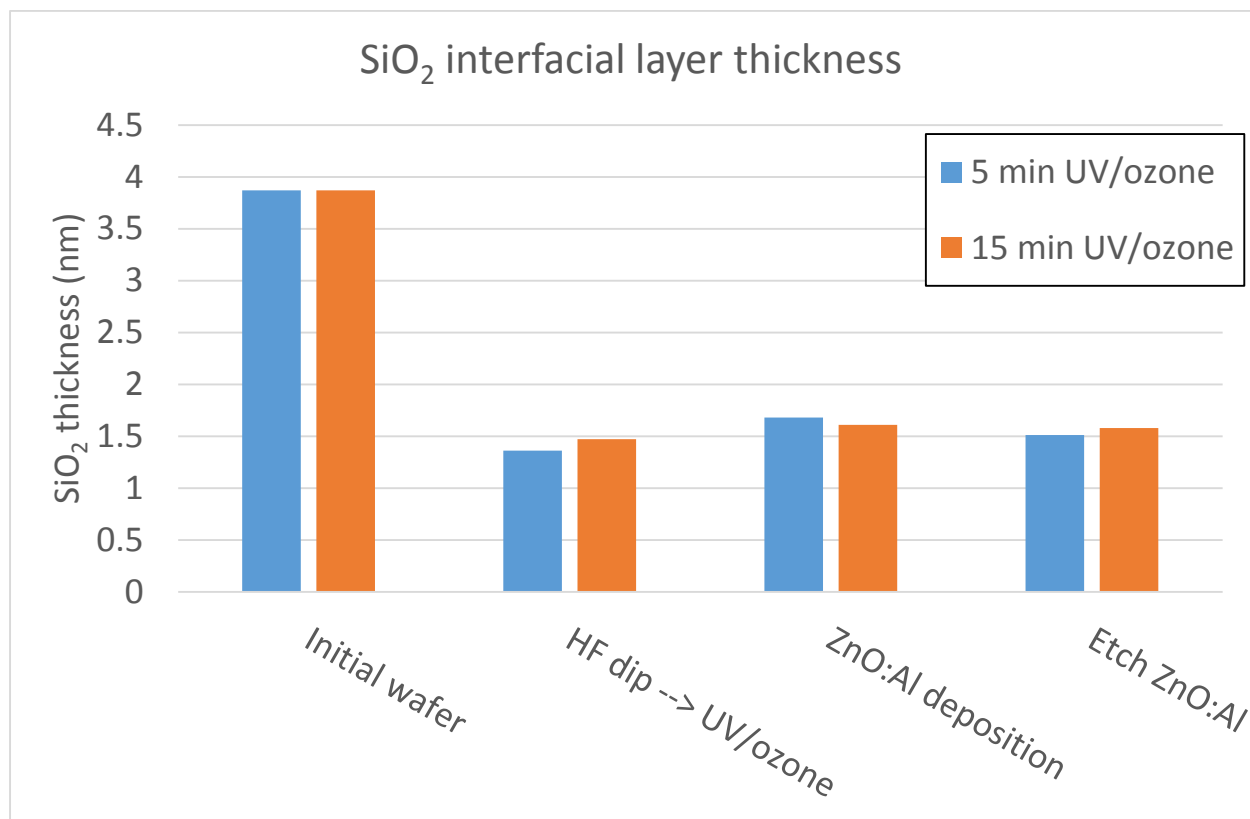

**Fig S5:** SiO$_2$ interfacial layer thickness, as measured by spectroscopic ellipsometry. The samples reported here used a 5 min UV/ozone oxidation step. The results of a 15 min exposure are also shown here for comparison.

4. Schottky barrier height measurement

    - We measured current density ($J$) as a function of voltage ($V$) and temperature ($T$) in the range 5 – 160 C. All measurements were performed in the dark. In **Figure S6** we show representative $J(V, T)$ data for pairs of *n*- and *p*-type diodes for one oxygen-poor and one oxygen-rich AZO film.
    - The ideality factor for our diodes is always well above unity, see Figure S6a (inset) and S6e (inset). This is consistent with the combined effects of the insulator layer in our M-I-S diodes and spatial inhomogeneity. Analyzing these effects is beyond the scope of this paper.
    - We determine the barrier heights $\Phi_p$ and $\Phi_n$ from Richardson plots of the reverse saturation current density, as shown in Figure S6d and S6h.

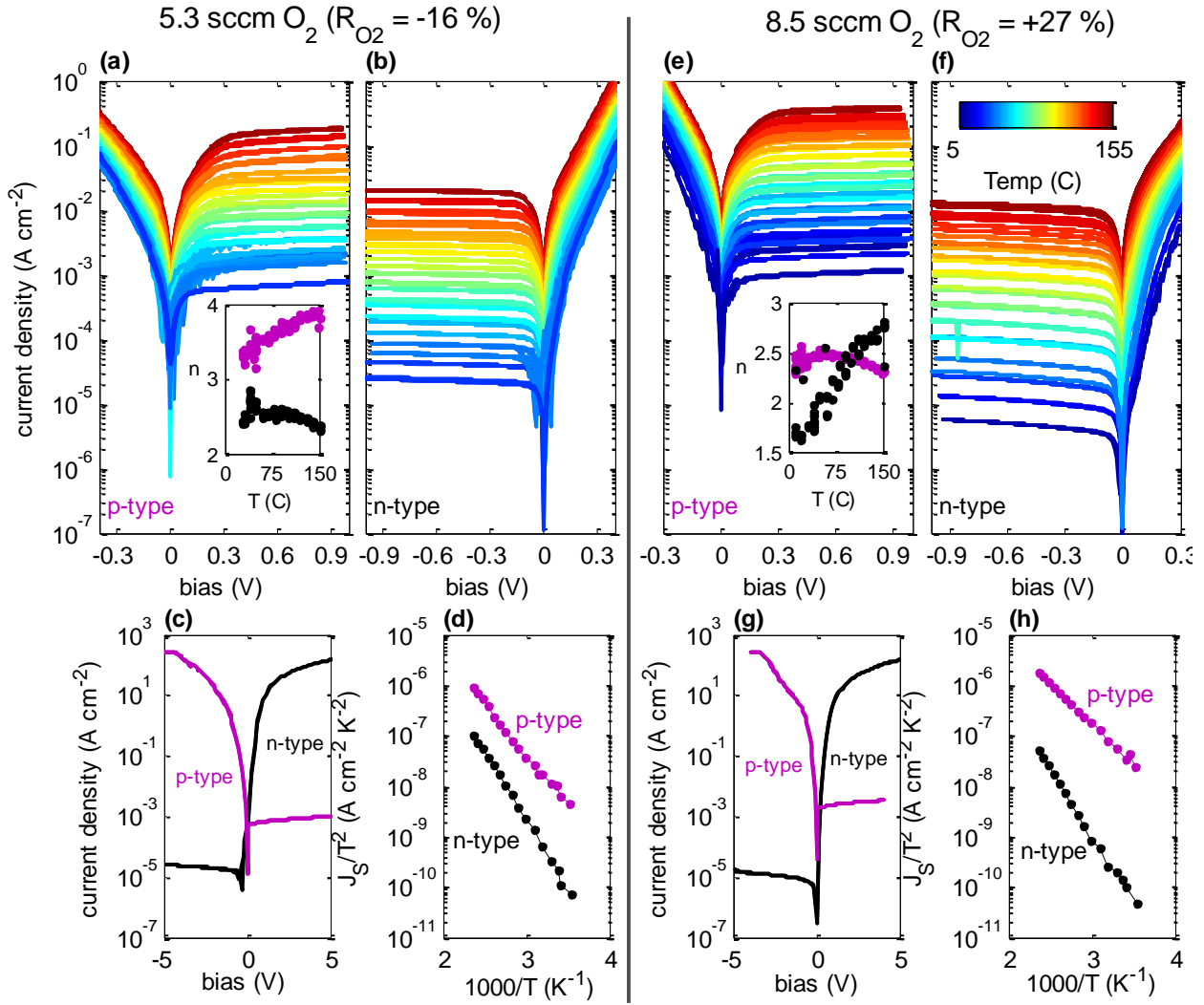

**Fig S6:** Measuring Schottky barrier heights. We show $J(V, T)$ data and Richardson plots for two AZO compositions, oxygen-poor ($R_{O2}$ = -16%, left half) and oxygen-rich ($R_{O2}$ = 27%, right half). **(a, e), (b, f)** Data for samples on *p*-type and *n*-type Si, respectively. The measurement $T$ is represented by the line color according to the color bar in (f). **(a & e, insets)** The diode ideality factors for *p*-type (purple) and *n*-type (black) samples. **(c, g)** Data at room-$T$ measured over a wider bias range. **(d, h)** The saturation current density ($J_S$) shown as a Richardson plot. The slope of these lines yields the barrier height for thermionic emission.

5. ZnO work function estimates
    - For the "electron affinity" data in Figure 2c we use a value of 4.5 eV for the work function of ZnO. The reported work function of ZnO and doped-ZnO (including AZO) varies over a wide range. We therefore use a simple average of all values reported in the following publications:
        - Gordon (2000).[8]
        - Sundaram & Kahn (1997). [13]
        - Gopel *et al.* (1980).[25]
        - Park *et al.* (2010).[26]
        - Schulze *et al.* (2007).[27]
        - Klein *et al.* (2009).[28]
        - Murdoch *et al.* (2009). [29]
        - Jiang *et al.* (2003).[30]
        - Kim *et al.* (2006).[31]
        - Chen *et al.* (2008).[9]
        - Huang *et al.* (2010).[32]
6. Barrier heights determined from data in Kim *et al.* (2001), ref. 15.
    - This paper reports on ZnO / Si photodiodes fabricated on lightly doped *n*- and *p*-type substrates. It does not report barrier height measurements. However, it does report reverse saturation current density for both *n*-type (see Table I) and *p*-type (see Figure 3) diodes. We use the ratio (*r*) of these values to determine $\Phi_p$-$\Phi_n$ using the equation $\Phi_p - \Phi_n = k_B T \ln(r)$.

- We note that, the exact value notwithstanding, our observation that the Fermi level sits in the upper half of the Si bandgap for undoped ZnO is consistent with the claims and discussion in ref. 15.

7. Atom probe sample preparation[33]
    - The sample preparation is illustrated by select electron micrographs in **Figure S7**
    - First we deposit a platinum protective layer (4×20 µm$^2$) until we have a layer of few hundred nanometers. Using ion beam milling we cut out a bar from the sample. We then lift out the bar by welding a micromanipulator to one of its free end. The bar is deposited onto flat tops by means of platinum welding. Each bar makes up to 6 APT samples. Each tip is sharpened using annular milling by gradually decreasing the size of the annulus and the beam current until desired radius and shank angle are reached. At the end, the tips go through a cleaning step at a 5 kV beam to reduce Ga implantation and damage from the Ga ion beam. Difference in contrast reveals different materials, which indicates that the interface AZO / SiO$_2$ / Si is captured within the needle-shaped tip.

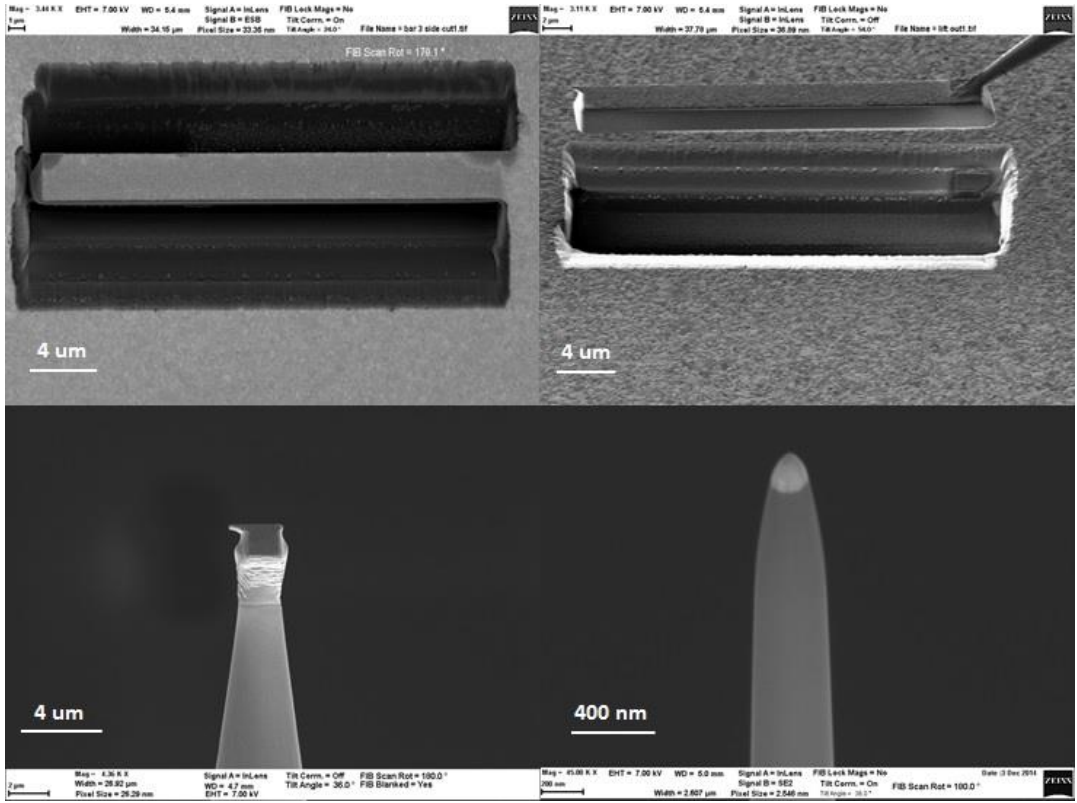

**Fig S7:** APT sample preparation. (**Top left**) Depositing protective layer and cutting a bar out of our sample. (**Top right**) Lift out step. (**Bottom left**) Deposited onto flat tops. (**Bottom right**) Annular milling and cleanup steps

8. Atom probe measurements and data analysis

    - The corrected mass spectrum is shown in **Figure S8**. The peaks are binned as shown, and the same binning was performed on all 6 samples to be able to cross-compare between them. The peak at 32 Da represent an overlap between $O_2^+$ and $Zn^{2+}$. Hence this peak was not used to interpret our results. The composition can still be obtained by using the remaining Zn isotopes labeled on the mass spectrum. Another misleading peak is the one at 16 Da which could be either $O_2^{2+}$ or $O^{1+}$, this is why the absolute oxygen concentration wasn't used in our analysis, and instead we scaled the cations concentration to the oxygen concentration. The peaks at 1, 2 and 3 Da correspond to hydrogen in the chamber, and do not affect the measurement and hence are not labeled. The peaks that have a hit count less than 1000 are not labeled here for clarity, but were taken into account for the analysis.
    - The elements labeled in the mass spectrum are reconstructed back to the position where they came from in the APT tip. The result is shown in **Figure S9** were only the listed elements are shown for better clarity of the AZO / $SiO_2$ / Si junction.
    - In the 3D reconstruction, the interface was defined as the depth in z, where we start collecting 5 at. % of Si. We then reduce our data to 1D concentration profiles of Al, Si, Zn and O calculated with respect to the defined interface with a bin size of 0.3 nm.
    - In **Figure S10** we plot proxigrams for Zn, O, Al, and Si concentrations (in at. %) for all six APT samples

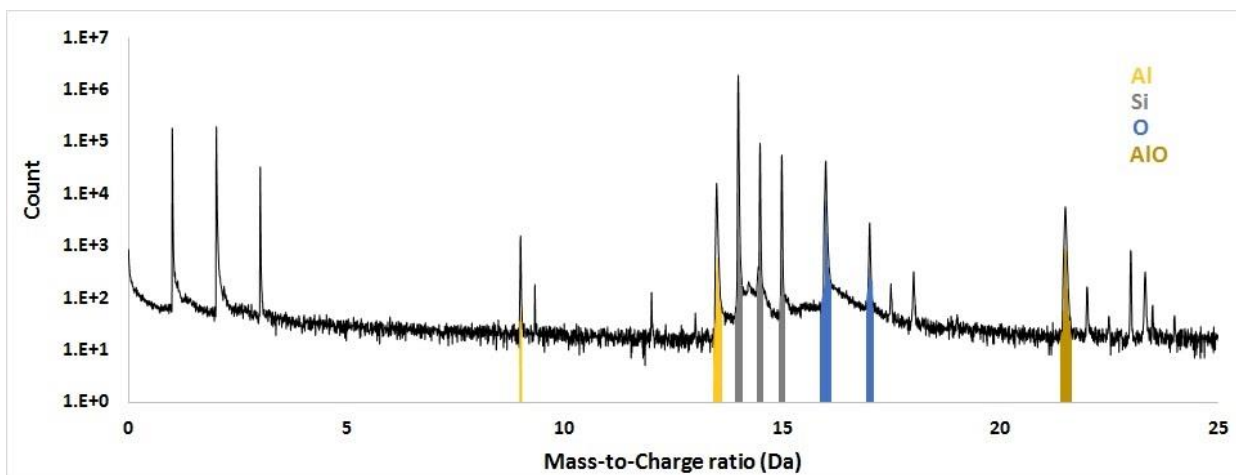
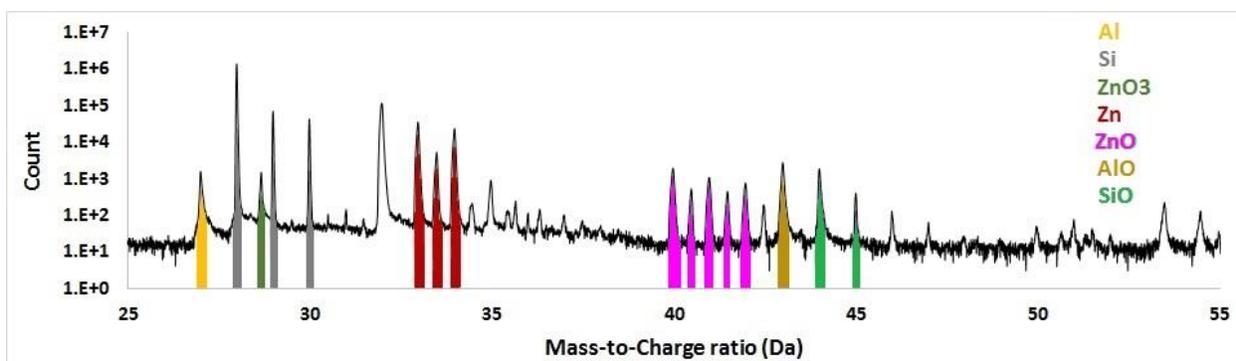
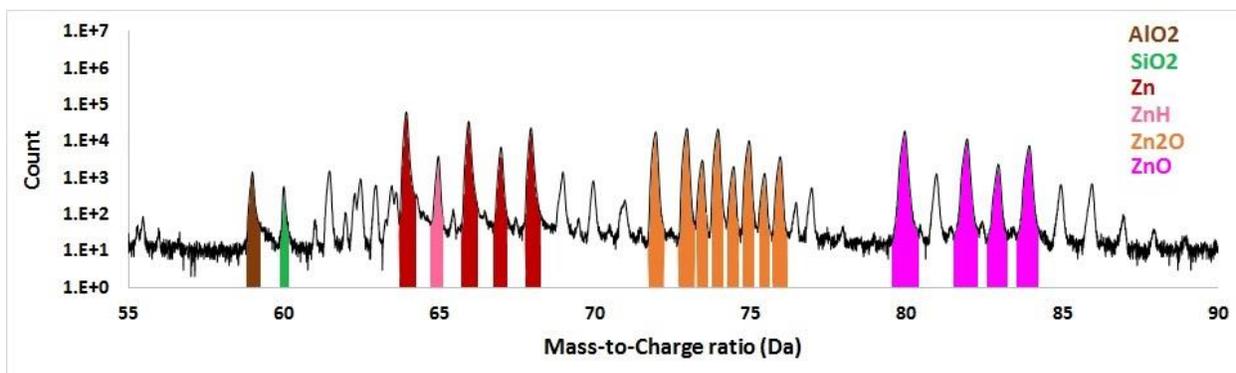

**Fig S8:** A representative mass spectrum divided into three sections for clearer interpretation.

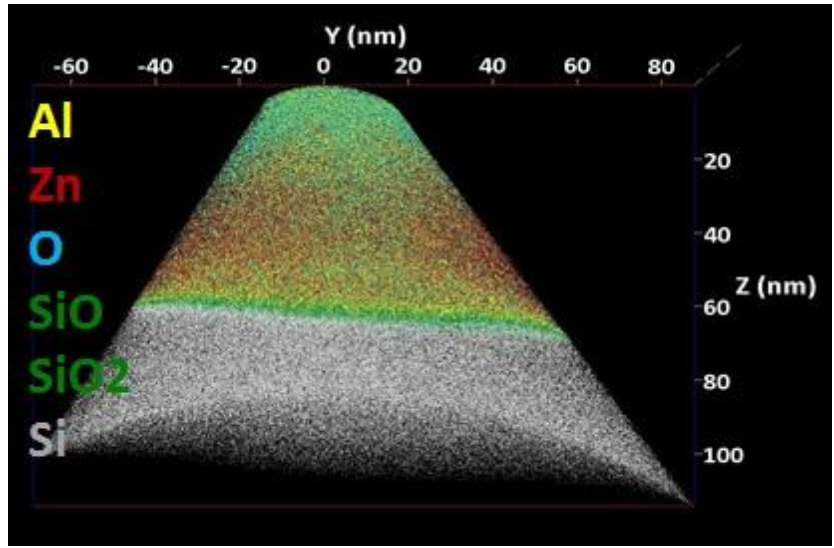

**Fig S9:** A 3D reconstruction of a representative APT tip showing the AZO film on top (approximately 60 nm thick) where the Al atoms are represented in yellow, Zn in red, O in blue, SiO and SiO2 in green and Si in grey.

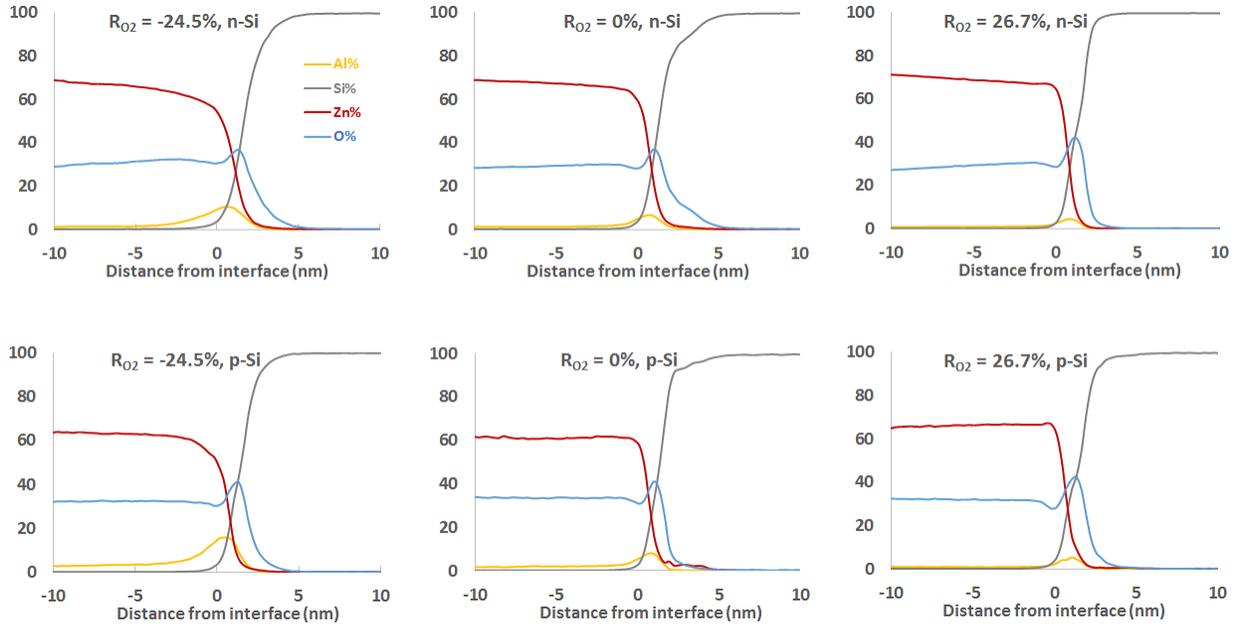

**Fig S10:** Proxigrams for Al, Zn, O and Si (at. %) for all 6 samples measured by APT. The concentration profiles are measured from the interface (5% Si) defined as zero on the x-axis.